# Advancing biological super-resolution microscopy through deep learning: a brief review

Tianjie Yang[1,2,3], Yaoru Luo[3,4], Wei Ji[1,2], Ge Yang[3,4,*]

1.Institute of Biophysics, Chinese Academy of Sciences, Beijing, 100101, China
2.College of Life Sciences, University of Chinese Academy of Sciences, Beijing, 100049, China
3.Laboratory of Computational Biology and Machine Intelligence, School of Artificial Intelligence, University of Chinese Academy of Sciences, Beijing, 100049, China
4.National Laboratory of Pattern Recognition, Institute of Automation, Chinese Academy of Sciences, Beijing, 100190, China
*. Corresponding Author: Ge Yang <ge.yang@ia.ac.cn>

**Abstract.** Super-resolution microscopy overcomes the diffraction limit of conventional light microscopy in spatial resolution. By providing novel spatial or spatiotemporal information on biological processes at nanometer resolution with molecular specificity, it plays an increasingly important role in life sciences. However, its technical limitations require trade-offs to balance its spatial resolution, temporal resolution, and light exposure of samples. Recently, deep learning has achieved breakthrough performance in many image processing and computer vision tasks. It has also shown great promise in pushing the performance envelope of super-resolution microscopy. In this brief Review, we survey recent advances in using deep learning to enhance performance of super-resolution microscopy. We focus primarily on how deep learning advances reconstruction of super-resolution images. Related key technical challenges are discussed. Despite the challenges, deep learning is set to play an indispensable and transformative role in the development of super-resolution microscopy. We conclude with an outlook on how deep learning could shape the future of this new generation of light microscopy technology.

**Keywords:** super-resolution microscopy, image super-resolution, fluorescence microscopy, deep learning, image reconstruction

# 1 Introduction

## 1.1 Introduction to biological super-resolution microscopy

Fluorescence microscopy is a light microscopy technology that plays a critical role in life sciences by capturing spatial or spatiotemporal information of biological processes [1, 2]. Its molecular specificity, low invasiveness, and multiplex capability make it a powerful tool for studying structure and function of biological processes in space and time at the molecular level under physiological conditions. However, the spatial resolution of conventional fluorescence microscopy is limited by the diffraction of visible light to ~200 nm. Under this resolution limit, many important molecular level details of biological processes are indistinguishable. Super-resolution microscopy overcomes this limit (Figure 1), routinely reaching spatial resolutions in the range of 20-70 nm [3-5], with some techniques reaching spatial resolutions of <10 nm in certain applications [6, 7]. Depending on their modes of image formation, the wide variety of super-resolution microscopy techniques generally fall under two categories: patterned illumination and single molecule localization.

Super-resolution by patterned illumination was pioneered by stimulated emission depletion (STED) microscopy (Figure 1C), which uses an intense doughnut-shaped depletion laser beam to create an emission region smaller than the diffraction limit [8, 9]. However, strong depletion illumination causes photobleaching and phototoxicity. A similar but more general photoswitching-based technique called reversible saturable optical linear fluorescence transitions (RESOLFT) was developed later, allowing substantially reduced depletion laser intensities [10]. Resolutions of these two techniques typically reach tens of nanometers. Under STED and RESOLFT, image acquisition requires point scanning of the field-of-view (FOV). In comparison, structured illumination microscopy (SIM) [11] is a widefield-based technique that uses patterned illumination to increase the spatial frequency that can be captured. It can use conventional fluorophores to image a large FOV on a millisecond timescale (Figure 1A) [12, 13]. However, it can only reach ~100 nm in spatial resolution. SIM using nonlinear illumination and photoswitchable proteins (NL-SIM) can further improve the resolution to ~50 nm [14], but to reach a higher resolution remains challenging. Under STED and RESOLFT, super-resolution images are directly acquired by point-scanning and photo-switching at defined spatial coordinates [3]. Under SIM and NL-SIM, super-resolution images are reconstructed through computational processing of acquired raw images.

Super-resolution by single molecule localization, often referred to as single molecule localization microscopy (SMLM), differentiates single switchable fluorophores within the diffraction limit by their blinking events. Stochastic optical reconstruction microscopy (STORM) (Figure 1E) [15] and photoactivation localization microscopy (PALM) [16] are two representatives. STORM uses switchable organic dyes while PALM uses photoactivatable proteins. SMLM can reach a spatial resolution of ~20-30 nm. However, blinking events under SMLM occur at random spatial coordinates.

Typically, thousands of raw images or more need to be collected so that enough localized single molecules can be accumulated to faithfully reconstruct the real geometry and fluorescence signal distributions of samples. For this reason, SMLM techniques require long acquisition time, and their low temporal resolution severely limits their applications in live cell imaging [4].

Super-resolution by patterned illumination and single molecule localization have complementary technical strengths. Recently, several techniques have been developed to combine these strengths by integrating the two strategies. In nanoscopy with minimal photon fluxes (MINFLUX) (Figure 1D), stochastic single molecule photoswitching is combined with patterned illumination-based localization to reach a resolution of ~1 nm [6]. In repetitive optical selective exposure (ROSE), multiple excitation illumination patterns are combined with stochastic single molecule photoswitching to reach a lateral resolution of ~5 nm [7] and an axial resolution of ~2 nm (Figure 1B) [17]. For these two techniques, enough single molecules must be localized for faithful reconstruction of super-resolution images, same as for STORM and PALM.

The super-resolution microscopy techniques introduced so far require specialized optics, specialized fluorophores, or both. Different from these techniques, several computational super-resolution techniques overcome the diffraction limit by analyzing random fluctuations of single fluorophore signals. Super-resolution optical fluctuation imaging (SOFI) [18] and super resolution radial fluctuations (SRRF) (Figure 1F)[19] are two representatives. They can be combined with other super-resolution techniques such as STORM and PALM or conventional widefield and confocal microscopy. Reconstruction of super-resolution images using these techniques requires computational processing of acquired raw images. Overall, along with the representative super-resolution microscopy techniques introduced here, many variants have been developed over the past two decades. See e.g., [20] for a case study. Comprehensive reviews of super-resolution microscopy techniques can be found in e.g., [3-5].

The wide variety of super-resolution microscopy techniques differ in their image formation and acquisition. However, from a computational point of view, there are important commonalities in their image reconstruction, which is the end goal of all super-resolution microscopy techniques. For the representative super-resolution microscopy techniques introduced here, Table 1 summarizes and compares their principles and performance goals of image reconstruction. The focus of this Review is on deep learning [21] based image reconstruction techniques for super-resolution microscopy.

Table 1. Comparison of image reconstruction for representative super-resolution techniques

| Modality | | Principle of Image Reconstruction | Main Performance Goals for Image Reconstruction |
|---|---|---|---|
| Image Formation | Representative Techniques | | |
| Patterned Illumination | STED, RESOLFT | Optical localization of single fluorophores at deterministic coordinates; Full frames acquired by point scanning | • To maximize spatial and temporal resolution<br>• To minimize light exposure |
| | SIM, NL-SIM | Full frames reconstructed from computational processing of raw widefield images | • To maximize spatial and temporal resolution<br>• To minimize reconstruction artifacts |
| Single molecule localization | STORM, PALM | Localization of single fluorophores through fitting at stochastic coordinates; Full frames reconstructed through data accumulation | • To maximize spatial and temporal resolution<br>• To minimize light exposure |
| Single molecule fluctuation | SOFI, SRRF | Localization of single fluorophores through fluctuation analysis; Full frames reconstructed from acquired raw image sequences | • To maximize spatial and temporal resolution<br>• To minimize reconstruction artifacts |
| Pattern Illumination + Single | | Localization of single fluorophores at stochastic coordinates; Full frames | • To maximize spatial and temporal resolution |

| Molecule Localization | MINFLUX, ROSE | reconstructed through data accumulation | • To minimize light exposure |

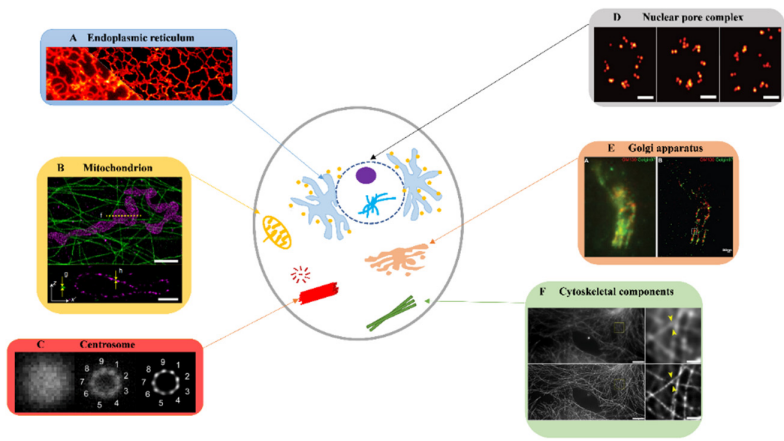

**Fig. 1.** Visualizing nanoscale structures of cells using different super-resolution microscopy techniques. (A) grazing incidence microscopy (left) and grazing incidence structured illumination microscopy (right) images of the endoplasmic reticulum in a live COS-7 cell. (B) Top: 3D two-color ROSE-Z images of α-tubulin and mitochondria outer membrane. Bottom: magnified view of the yellow line scan. (C) confocal microscopy, STED and deconvolved STED images of centrosomal protein asterless in a D. Mel-2 cell. (D) MINFLUX nanoscopy images of Nucleoporin (Nup96) of nuclear pore complexes in a U2OS cell. (E) Cis-Golgi marker (GM130) and trans-Golgi marker (Golgin-97) visualized by conventional fluorescence microscope (left) and two-color STORM (right) in a COS-7 cell. (F) widefield (top) and SRRF (bottom) reconstruction images of microtubules in a fixed cell. Images are adapted with permission from the following sources: A: [12], B: [17], C: [121], D: [6], E: [122], F: [19].

## 1.2    Introduction to deep learning for image processing and computer vision

Deep learning refers to a class of machine learning or artificial intelligence techniques that compute using artificial neural networks with many layers, often called deep neural networks (DNNs) [21]. Honored by the 2018 ACM Turing Award, it has revolutionized how we analyze and understand images and has been used with tremendous success in virtually all kinds of image processing and computer vision tasks, such as image classification [22], object detection [23, 24], image segmentation [25], object tracking [26, 27], image registration [28, 29], image denoising [30], and image synthesis [31, 32]. The most commonly used types of DNNs for such tasks are convolutional neural networks [21] and, for image sequences, recursive neural networks [33].

Other types of neural networks such as graph neural networks [34, 35] have also been used for various applications. Most of the deep learning-based image processing and computer vision techniques are developed for natural images.

The first step in solving an image processing or computer vision problem using deep learning typically is to decide on a learning strategy, such as supervised learning, semi-supervised learning, or unsupervised learning [36-38]. The decision is usually based on the availability of labeled training data and the cost of producing new labeled training data. However, training with unlabeled data can help prevent overfitting [36-38]. DNNs are trained with fully labelled data in supervised learning, partially labelled data in semi-supervised learning, and unlabeled data in unsupervised learning.

The next step is to choose an existing DNN architecture or to develop a new or custom DNN architecture, also referred to as a model, for the best performance. Indeed, many DNN architectures have been developed over the past decade (Figure 2). In image classification, for example, the ImageNet Large Scale Visual Recognition Challenge (ILSVRC) [39] has played a particularly important role in driving the development of new models and in starting the deep learning revolution. Representative models coming out of this competition include AlexNet [40], VGG [41], Inception (GoogLeNet) [42], and ResNet (Figure 2A) [43], to name a few. In object detection, image objects are located and classified by assigning rectangular bounding boxes. One-stage detectors refer to models that combine localization and classification into one step. Representative one-stage detectors include YOLO [44], SSD [45], RetinaNet [46] and CornerNet [47]. Two-stage detectors separate object localization and classification into two steps. Representative two-stage detectors include R-CNN [48], Fast R-CNN [49], Faster R-CNN (Figure 2C) [50], R-FCN [51] and Mask R-CNN [52]. In semantic image segmentation, individual pixels belonging to the same object are grouped and assigned the same semantic label. Representative models include fully convolutional network (FCN) [53], SegNet [54], U-Net (Figure 2B) [55]. More recent segmentation models use multiscale image features. Representative models include Pyramid Scene Parsing Network (PSPN) [56], Adaptive Pyramid Context Network (APC-Net) [57], Multi-Scale Context Intertwining (MSCI) [58] and High-Resolution Network (HRNet) [59]. Most recently, the Transformer architecture [60, 61], originally developed for natural language processing, has found substantial success in image processing and computer vision tasks such as image classification [62, 63].

The selection or development of a DNN model is usually accompanied by the selection or development of an application-oriented loss function [64, 65]. Training of DNNs is essentially a process that optimizes their connection weights to minimize or maximize the loss function, called the cost function in optimization. Various optimization methods can be used for the training [66, 67]. Configuration parameters used in training of DNNs, called hyperparameters, require tuning [68, 69]. Overall, in assessing a deep learning technique, key components to be considered include its learning strategy, network architecture, loss function, and training data. Its optimization method and hyperparameters are key to its implementation.

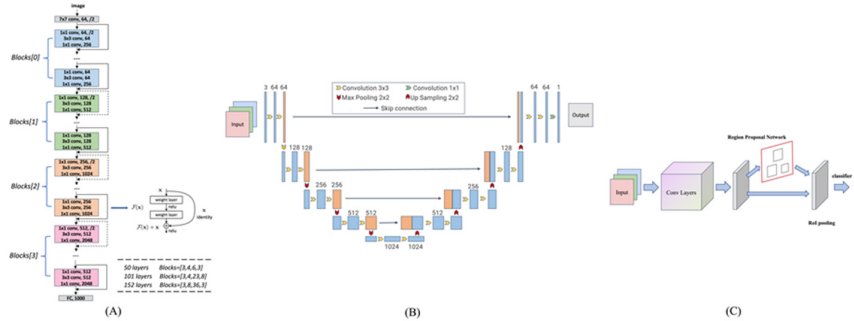

**Fig. 2.** Representative architectures of deep neural networks. (A) ResNet: a feedforward architecture with residual blocks, often used for image classification. (B) U-Net: an encoder-decoder architecture with skip connections at different scales, often used for image segmentation. (C) Faster R-CNN: a two-stage detector architecture, often used for object detection.

### 1.3 Introduction to deep learning for processing fluorescence microscopy images

In addition to its tremendous success in processing natural images, deep learning has also found great success in processing fluorescence microscopy images [70-73]. From a user's perspective, deep learning techniques offers at least two important advantages over traditional image processing techniques. First, they offer superior performance. A striking example is the synthesis of realistic fluorescence microscopy images [74] using generative adversarial networks (GANs) [32]. It is not feasible for traditional image synthesis techniques to achieve the same level of fidelity. Second, deep learning techniques are more user friendly. Once DNNs are trained properly, they can be used without parameter tuning. In contrast, parameter tuning is often essential for traditional image processing techniques.

Despite the great success of deep learning in processing both natural images and fluorescence microscopy images, it is important to note the differences between these two types of images. First, fluorescence microscopy images have simpler structures and semantics than natural images. For each color (i.e. wavelength) channel, the semantic label for each pixel is binary, either foreground or background, and the image background is composed of structure-free regions of noise. Second, fluorescence microscopy images have greater pixel depths and wider dynamic ranges than natural images. Third, fluorescence microscopy images have noise properties that differ substantially from those of natural images [75]. Fourth, blurring of image objects is common in fluorescence microscopy images because of limited depth-of-field. Overall, it is essential to consider these distinct properties in developing deep learning-based processing techniques for fluorescence microscopy images. It is also essential to consider the technical limitations of deep learning techniques, which are discussed in Section 3.

## 1.4 Organization and aim

This Review is organized as follows: Section 1 provides necessary background information on biological super-resolution microscopy and deep learning. Section 2 surveys representative works in using deep learning techniques to advance reconstruction of super-resolution microscopy images. Section 3 concludes with a discussion of key technical challenges and an outlook on how deep learning could shape the future of super-resolution microscopy. Overall, this Review aims to provide a concise, in-depth, and up-to-date survey of related works for researchers and practitioners interested in super-resolution microscopy and deep learning.

## 2 Deep learning for reconstruction of super-resolution microscopy images

Super-resolution microscopy techniques are defined by their capability to overcome the diffraction limit in spatial resolution. However, spatial resolution is not the only performance metric required by real-world applications. Other important performance metrics include temporal resolution, length of image acquisition, and area of image acquisition (i.e. FOV), etc. In addition to diverse performance requirements, super-resolution microscopy techniques are subject to various constraints that define their performance envelopes. For example, SMLM techniques such as STORM and PALM are limited in their temporal resolutions. Samples are limited in the amount of photons they can tolerate before photobleaching and photodamage [76]. Overall, for super-resolution microscopy techniques, trade-offs must be made between performance metrics within their performance envelopes to meet requirements of different applications. Deep learning techniques provide a potentially transformative solution to enhance performance of super-resolution microscopy techniques and to push their performance envelopes [73]. Because image reconstruction is at the core of all super-resolution microscopy techniques, we focus on examining recent advances in using deep learning to advance reconstruction of super-resolution microscopy images.

### 2.1 Deep learning-based enhancement of spatial resolutions

Interestingly, the term "super-resolution" was first coined for natural images and is defined as overcoming resolution limits of optical imaging systems by image processing [77-79]. This definition certainly is also valid for biological super-resolution microscopy. However, unlike in biological super-resolution microscopy, what resolution qualifies as "super-resolution" for natural images is more loosely defined and is often judged by human perception. Research on this topic has a long history and dates back at least to 1980s [79]. Conventional super-resolution techniques developed before the rise of deep learning techniques are examined in several reviews [77-79]. However, it is deep learning that has enabled transformative performance advances [80, 81]. A wide variety of deep learning-based super resolution techniques have been developed for natural images [80-82]. Among them, single-image super-resolution techniques that require just one input image have been extensively studied [82-84] and have recently been used for enhancing the resolution of fluorescence microscopy [85, 86].

Overall, the implementation of deep learning-based super-resolution techniques for both natural images and fluorescence microscopy images follow the same supervised learning scheme that consists of two steps: training data preparation and model training. In training data preparation, paired and aligned low-resolution and high-resolution images of the same FOV are produced by experiments, computer simulation, or a mixture of both. In model training, the paired images are used to train DNNs to learn the mapping between the low-resolution image domain as the input domain and the high-resolution image domain as the output domain. After the models are trained, they are used to transform an input of low-resolution images into an output of high-resolution images. In this way, deep learning enables computational reconstruction of synthetic high-resolution images. This process is presumably more convenient and cost-effective than physical acquisition of real high-resolution images. Specifically for fluorescence microscopy images, deep learning based super-resolution techniques provide a computational solution that is potentially transformative in enhancing spatial resolutions and overcoming the diffraction limit if properly implemented.

To date, a significant number of studies have reported using deep learning to enhance spatial resolutions in image reconstruction for biological super-resolution microscopy. Overall there are two application scenarios. Under the first scenario, deep learning is used directly to reconstruct super-resolution microscopy images of higher resolutions. For example, Nehme and colleagues have reported using DNNs to enhance the performance of single molecule localization for reconstruction of STORM images in 2D and 3D [87, 88]. Deep learning achieves higher localization accuracy than conventional point spread function (PSF) fitting under high fluorophore density and low SNR with real-time speed and no parameter tuning for 2D STORM [88]. They extend their work to 3D STORM by combining PSF engineering with deep learning-based single molecule localization and PSF pattern recognition [87]. Indeed, deep learning is well suited for recognition of complex patterns of engineered PSFs and has achieved superior detection accuracy and speed in image reconstruction in several other studies, e.g., [89, 90]. Under the second scenario, deep learning is used to computationally reconstruct a high-resolution image from a low-resolution image. For example, Wang and colleagues use a GAN network to transform low-resolution images into high-resolution images across different modalities, such as from low-NA widefield to high-NA widefield, from confocal to STED, and from conventional TIRF to TIRF-SIM [86]. Fang and colleagues use a U-Net type model to enhance resolutions of electron microscopy images and fluorescence microscopy images [85]. Qiao and colleagues use two deep learning models for enhancing performance of SIM under low signal-to-noise-ratios (SNRs) and long intervals of imaging [91].

Key components of reviewed studies under the two scenarios are summarized and compared in Table 2. Several observations can be made from the comparison. First, if paired training data is available, deep learning can be used to enhance resolution across different modalities, including widefield microscopy, confocal microscopy, SMLM, SIM, and transmission electron microscopy, demonstrating the versality of the approach. The training data may be produced by a mixture of simulation and experiments or entirely by simulation. Overall, generalization capability and robustness of the proposed models are not thoroughly characterized in the reviewed studies. Second, performance metrics used for super-resolution microscopy images are similar as those used for natural images, such as PSNR (peak signal-to-noise ratio) and SSIM (structural similarity index). These metrics may not be well suited for fluorescence microscopy images because of their differences from natural images. Third, generation of artifacts has been reported in all the studies reviewed. Currently there is no systematic solution to this problem.

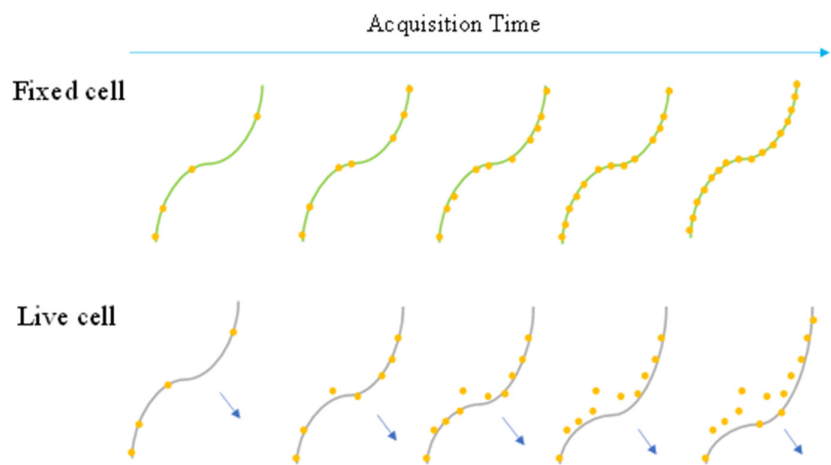

**Fig. 3.** A cartoon illustration of single molecule localization on a curvilinear structure in a fixed cell versus a live cell. In the fixed cell, localized single molecule coordinates are samples of the same static underlying structure. In the live cell, localized single molecule coordinates are samples of the varying underlying structure at different time points.

## 2.2  Deep learning-based noise reduction

Because of the basic principle of their image formation, SMLM techniques require long image acquisition, which severely limits their applications in live cell imaging. Increasing fluorescence labeling density can accelerate SMLM, but it will also cause overlap of single fluorophore signals in the diffraction limit. Nehme and colleagues have shown that deep learning is capable of localizing single fluorophores at higher density [87, 88]. Still, the allowed density has a limit. In addition to increasing labeling density, a variety of chemical and physical strategies have been proposed to overceome the limitation in temporal resolution. Brighter fluorophores, stronger laser and faster cameras have all been tried to reduce acquisition time [92-94]. However, fast acquisition may reduce image quality while strong laser may induce photobleaching and photodamage. Overall, none of these strategies address the fundamental constraint of SMLM, namely large numbers of single molecules must be localized for faithful image reconstruction. To this end, several studies have tried to reduce the number of required localized single molecules using compressed sensing [95, 96] and sparse support [97]. Recently, DNNs have demonstrated superior performance and great potential in reconstructing SMLM images from sparse data [98, 99]. Ouyang and colleagues use the pix2pix GAN to generate high-resolution SMLM images from sparse images of localized single molecules [98]. In comparison, Gaire and colleagues use a much simpler residual learning architecture for reconstructing high-resolution SMLM images from sparse data for up to three color channels [99].

So far, we have focused on deep learning for reconstruction of single super-resolution images from sparse data. In live cell imaging, acquired videos has substantial spatial and temporal continuity. Deep learning has been demonstrated to extract information in videos to enhance spatial and temporal resolution of confocal and light-field microscopy [85, 100]. However, specifically for SMLM, deep learning faces new challenges in assigning localized single fluorophores to moving structures (Figure 3)

### 2.3 Other applications of deep learning

SNR is an important performance metric for fluorescence microscopy. Low SNR images are difficult to analyze. Increasing laser power is a direct way to improve the SNR but will also increase the likelihood of photobleaching and photodamage. Classical denoising algorithms has been used to enhance SNR of fluorescence images under low laser power [101-103]. Deep learning-based methods have been shown to have superior performance over classical methods in recent studies [104]. So far, deep learning-based denoising has been used in image acquisition of several super-resolution microscopy modalities, including SIM [105], STED [106], and SMLM [107]. It is expected that deep learning based denoising will be widely adopted in super-resolution microscopy to increase SNR and reduce photodamage of biology sample.

### 2.4 Other applications of deep learning

Multicolor super-resolution microscopy uses different fluorescence probes to reveal multiple molecular structure at same time. However fast multicolor imaging leads to overlap of emission spectra which causes channel mixing and reduces the final resolution. Deep learning-based methods are used in separating mixing spectra during image acquisition in recent studies [92, 99, 108]. These strategies have shown excellent performance in decreasing cross-color contamination, accelerating image acquisition, and increasing the final resolution of super-resolution microscopy. In addition, deep learning has found many other applications in fluorescence microscopy. So far, a common strategy for 3D fluorescence microscopy is to use specialized optics and engineered PSF. A recent study has shown that deep CNN can extract information from single 2D image to reconstruct a 3D filed of the sample [109]. This may bring in a new approach for 3D super-resolution microscopy.

**Table 2.** Representative studies using deep learning to enhance spatial and temporal resolution of super-resolution microscopy

| Application | Model | Imaging Modality | Network architecture | Loss Function | Training Data | Performance Metric | Reference |
|---|---|---|---|---|---|---|---|
| Enhancement of spatial resolution | Deep-STORM | 2D-STORM | Deep-STORM | MSE with L1 regularization | Synthetic & Experimental | NMSE | [88] |
| | | 3D-STORM with engineered PSF | Deep-STORM3D | MSE with L1 regularization | Synthetic | NMSE | [87] |
| | smNET | 2D-SMLM with engineered PSF | Customized ResNet | MSE | Synthetic & experimental | MSE | [90] |
| | -- | 3D-STORM with engineered PSF | VGG16 | MSE | Synthetic | MSE | [89] |

|  | | | | | | |
|---|---|---|---|---|---|---|
| -- | Widefield, point-scan confocal, TIRF | GAN (U-Net) | MSE, SSIM | Paired, experimental | FWHM | [86] |
| -- | SEM, point-scan confocal | Res-UNet | MSE | Paired, synthetic | PSNR, SSIM, FRC, NanoJ-Squirrel | [85] |
| DFCAN, DFGAN | SIM | DFCAN DFGAN | MSE, SSIM (DFCAN) MSE, SSIM, (DFGAN) | Paired, experimental (with augmentation | NRMSE, MS-SSIM, decorrelation | [91] |
| Enhancement of temporal resolution | | | | | | |
| ANNA-PALM | PALM | pix2pix (GAN) | MSE, MS-SSIM | Paired, experimental & synthetic | MS-SSIM | [98] |
| -- | Spectroscopic SMLM | Customized ResNet | MSE | Experimental | FWHM, MS-SSIM | [99] |

## 3  Discussion and outlook

Today, models with millions parameters are common in deep learning [110]. In a somewhat extreme case, the GPT-3 model for natural language processing contains 175 billon parameters [111]. The enormous numbers of parameters, which far exceed those of traditional image process and computer vision algorithms, are one of the key factors that give DNNs the power to handle challenging image processing and computer vision tasks [112, 113]. Indeed, as demonstrated by the works reviewed here, deep learning has great potential in pushing the performance envelope of super-resolution microscopy. In the meantime, however, it also faces critical technical challenges. First, to minimize artifacts in reconstruction of super-resolution images is a key challenge for deep learning-based performance enhancement techniques [114]. So far, although progress has been made [115], characterization and minimization of artifacts remain an open problem. Second, to ensure generalization of DNNs for super-resolution microscopy is also a key challenge. DNNs trained on images collected on selected microscopes under selected conditions may not perform well on images collected on other microscopes under other conditions [116]. Third, to ensure robustness of DNNs for super-resolution microscopy is another key challenge. Fluctuations in imaging conditions are common in fluorescence microscopy, especially in live cell imaging. Deep-learning algorithms are data driven and are known to be sensitive to such fluctuations. In fact, if not trained properly, performance of DNNs can collapse under such fluctuations [117]. Lastly, to ensure interpretability of DNNs for super-resolution microscopy is also a key challenge in real-world applications [118]. Currently, deep learning lacks a rigorous theoretical foundation [119] for in-depth understanding of its basic properties such as generalization, robustness, and interpretability.

Generation of artifacts is a common issue in solving inverse problems such as image reconstruction [120]. Looking into the future, we expect that this issue will be gradually resolved through the synergy of multiple measures, including but not limited to rigorous experimental control, incorporation of realistic physical models, and improvement in design of DNN architectures and loss functions. The challenges of ensuring generalization, robustness, and interpretability of DNNs are universal for all deep learning applications, not just super-resolution microscopy. They will be gradually overcome by advances in the general theory and practice of deep learning as well as customized solutions for super-resolution microscopy. Overall, despite the challenges, we believe deep learning is set to play an indispensable and transformative role in advancing super-resolution microscopy as the next generation of light microscopy technology.

**Acknowledgements.** The work was supported in part by the Strategic Priority Research Program of Chinese Academy of Sciences (grant XDB37040402 to G.Y. and grant XDB37040104 to W.J.), the National Natural Science Foundation of China grant 91954201 under the major research program "Organellar interactomes for cellular homeostasis" and grant 31971289 to G.Y.), the Chinese Academy of Sciences

(grant 292019000056), and the University of Chinese Academy of Sciences (grant 115200M001).

**References**

1. Lichtman, J.W. and J.-A. Conchello, *Fluorescence microscopy.* Nature Methods, 2005. **2**(12): p. 910-919.
2. Giepmans, B.N.G., et al., *The fluorescent toolbox for assessing protein location and function.* Science, 2006. **312**(5771): p. 217-224.
3. Sahl, S.J., S.W. Hell, and S. Jakobs, *Fluorescence nanoscopy in cell biology.* Nature Reviews Molecular Cell Biology, 2017. **18**(11): p. 685-701.
4. Sigal, Y.M., R. Zhou, and X. Zhuang, *Visualizing and discovering cellular structures with super-resolution microscopy.* Science, 2018. **361**(6405): p. 880-887.
5. Valli, J., et al., *Seeing beyond the limit: a guide to choosing the right super resolution microscopy technique.* Journal of Biological Chemistry, 2021: p. 100791.
6. Balzarotti, F., et al., *Nanometer resolution imaging and tracking of fluorescent molecules with minimal photon fluxes.* Science, 2017. **355**(6325): p. 606-612.
7. Gu, L., et al., *Molecular resolution imaging by repetitive optical selective exposure.* Nature Methods, 2019. **16**(11): p. 1114-1118.
8. Hell, S.W. and J. Wichmann, *Breaking the diffraction resolution limit by stimulated-emission - stimulated-emission-depletion fluorescence microscopy.* Optics Letters, 1994. **19**(11): p. 780-782.
9. Klar, T.A., et al., *Fluorescence microscopy with diffraction resolution barrier broken by stimulated emission.* Proceedings of the National Academy of Sciences, 2000. **97**(15): p. 8206-8210.
10. Hofmann, M., et al., *Breaking the diffraction barrier in fluorescence microscopy at low light intensities by using reversibly photoswitchable proteins.* Proceedings of the National Academy of Sciences, 2005. **102**(49): p. 17565-17569.
11. Gustafsson, M.G.L., *Surpassing the lateral resolution limit by a factor of two using structured illumination microscopy.* Journal of Microscopy, 2000. **198**(2): p. 82-87.
12. Guo, Y.T., et al., *Visualizing intracellular organelle and cytoskeletal interactions at nanoscale resolution on millisecond timescales.* Cell, 2018. **175**(5): p. 1430-1442.

13. Li, D., et al., *Extended-resolution structured illumination imaging of endocytic and cytoskeletal dynamics.* Science, 2015. **349**(6251): p. aab3500.

14. Rego, E.H., et al., *Nonlinear structured-illumination microscopy with a photoswitchable protein reveals cellular structures at 50-nm resolution.* Proceedings of the National Academy of Sciences, 2012. **109**(3): p. E135-E143.

15. Rust, M.J., M. Bates, and X.W. Zhuang, *Sub-diffraction-limit imaging by stochastic optical reconstruction microscopy (STORM).* Nature Methods, 2006. **3**(10): p. 793-795.

16. Betzig, E., et al., *Imaging intracellular fluorescent proteins at nanometer resolution.* Science, 2006. **313**(5793): p. 1642-1645.

17. Gu, L., et al., *Molecular-scale axial localization by repetitive optical selective exposure.* Nature Methods, 2021. **18**(4): p. 369-373.

18. Dertinger, T., et al., *Fast, background-free, 3D super-resolution optical fluctuation imaging (SOFI).* Proceedings of the National Academy of Sciences, 2009. **106**(52): p. 22287-22292.

19. Gustafsson, N., et al., *Fast live-cell conventional fluorophore nanoscopy with ImageJ through super-resolution radial fluctuations.* Nature Communications, 2016. **7**(1): p. 12471.

20. Sage, D., et al., *Super-resolution fight club: assessment of 2D and 3D single-molecule localization microscopy software.* Nature Methods, 2019. **16**(5): p. 387-395.

21. LeCun, Y., Y. Bengio, and G. Hinton, *Deep learning.* Nature, 2015. **521**(7553): p. 436-444.

22. Rawat, W. and Z. Wang, *Deep convolutional neural networks for image classification: a comprehensive review.* Neural Computation, 2017. **29**(9): p. 2352-2449.

23. Zhao, Z., et al., *Object detection with deep learning: a review.* IEEE Transactions on Neural Networks and Learning Systems, 2019. **30**(11): p. 3212-3232.

24. Liu, L., et al., *Deep learning for generic object detection: A survey.* International Journal of Computer Vision, 2020. **128**(2): p. 261-318.

25. Garcia-Garcia, A., et al., *A review on deep learning techniques applied to semantic segmentation.* arXiv preprint arXiv:1704.06857, 2017.

26. Ciaparrone, G., et al., *Deep learning in video multi-object tracking: a survey.* Neurocomputing, 2020. **381**: p. 61-88.

27. Li, P., et al., *Deep visual tracking: Review and experimental comparison.* Pattern Recognition, 2018. **76**: p. 323-338.

28. Fu, Y., et al., *Deep learning in medical image registration: a review.* Physics in Medicine & Biology, 2020. **65**(20): p. 20TR01.

29. Haskins, G., U. Kruger, and P. Yan, *Deep learning in medical image registration: a survey.* Machine Vision and Applications, 2020. **31**(1): p. 8.

30. Tian, C., et al., *Deep learning on image denoising: An overview.* Neural Networks, 2020. **131**: p. 251-275.

31. Shorten, C. and T.M. Khoshgoftaar, *A survey on image data augmentation for deep learning.* Journal of Big Data, 2019. **6**(1): p. 60.

32. Wang, Z., Q. She, and T. Ward, *Generative adversarial networks: a survey and taxonomy.* arXiv preprint arXiv:1906.01529, 2019.

33. Yu, Y., et al., *A review of recurrent neural networks: LSTM cells and network architectures.* Neural Computation, 2019. **31**(7): p. 1235-1270.

34. Zhou, J., et al., *Graph neural networks: A review of methods and applications.* AI Open, 2020. **1**: p. 57-81.

35. Wu, Z., et al., *A comprehensive survey on graph neural networks.* IEEE Transactions on Neural Networks and Learning Systems, 2021. **32**: p. 4-24.

36. van Engelen, J.E. and H.H. Hoos, *A survey on semi-supervised learning.* Machine Learning, 2020. **109**(2): p. 373-440.

37. Schmarje, L., et al., *A survey on semi-, self- and unsupervised learning for image classification.* IEEE Access, 2021. **9**: p. 82146-82168.

38. Karhunen, J., T. Raiko, and K. Cho, *Chapter 7 - Unsupervised deep learning: A short review*, in *Advances in Independent Component Analysis and Learning Machines*, E. Bingham, et al., Editors. 2015, Academic Press. p. 125-142.

39. Russakovsky, O., et al., *Imagenet large scale visual recognition challenge.* International Journal of Computer Vision, 2015. **115**(3): p. 211-252.

40. Krizhevsky, A., I. Sutskever, and G.E. Hinton, *Imagenet classification with deep convolutional neural networks.* Advances in Neural Information Processing Systems, 2012. **25**: p. 1097-1105.

41. Simonyan, K. and A. Zisserman, *Very deep convolutional networks for large-scale image recognition.* arXiv preprint arXiv:1409.1556, 2014.

42. Szegedy, C., et al. *Going deeper with convolutions*. in *Proceedings of the IEEE Conference on Computer Vision and Pattern Recognition (CVPR)*. 2015.

43. He, K., et al. *Deep residual learning for image recognition*. in *Proceedings of the IEEE Conference on Computer Vision and Pattern Recognition (CVPR)*. 2016.

44. Redmon, J., et al. *You only look once: Unified, real-time object detection*. in *Proceedings of the IEEE Conference on Computer Vision and Pattern Recognition (CVPR)*. 2016.

45. Liu, W., et al. *Ssd: Single shot multibox detector*. in *Proceedings of the European Conference on Computer Vision (ECCV)*. 2016. Springer.

46. Lin, T.-Y., et al. *Focal loss for dense object detection*. in *Proceedings of the IEEE International Conference on Computer Vision (ICCV)*. 2017.

47. Law, H. and J. Deng. *Cornernet: Detecting objects as paired keypoints*. in *Proceedings of the European Conference on Computer Vision (ECCV)*. 2018.

48. Girshick, R., et al. *Rich feature hierarchies for accurate object detection and semantic segmentation*. in *Proceedings of the IEEE Conference on Computer Vision and Pattern Recognition (CVPR)*. 2014.

49. Girshick, R. *Fast R-CNN*. in *Proceedings of the IEEE International Conference on Computer Vision (ICCV)*. 2015.

50. Ren, S., et al., *Faster R-CNN: Towards real-time object detection with region proposal networks.* arXiv preprint arXiv:1506.01497, 2015.

51. Dai, J., et al., *R-FCN: Object detection via region-based fully convolutional networks.* arXiv preprint arXiv:1605.06409, 2016.

52. He, K., et al. *Mask R-CNN*. in *Proceedings of the IEEE International Conference on Computer Vision (ICCV)*. 2017.

53. Long, J., E. Shelhamer, and T. Darrell. *Fully convolutional networks for semantic segmentation*. in *Proceedings of the IEEE Conference on Computer Vision and Pattern Recognition (CVPR)*. 2015.

54. Badrinarayanan, V., A. Kendall, and R. Cipolla, *Segnet: A deep convolutional encoder-decoder architecture for image segmentation.* IEEE Transactions on Pattern Analysis and Machine Intelligence, 2017. **39**(12): p. 2481-2495.

55. Ronneberger, O., P. Fischer, and T. Brox. *U-net: Convolutional networks for biomedical image segmentation*. in *International Conference on Medical Image Computing and Computer-Assisted Intervention (MICCAI)*. 2015. Springer.

56. Zhao, H., et al. *Pyramid scene parsing network*. in *Proceedings of the IEEE Conference on Computer Vision and Pattern Recognition (CVPR)*. 2017.

57. He, J., et al. *Adaptive pyramid context network for semantic segmentation*. in *Proceedings of the IEEE/CVF Conference on Computer Vision and Pattern Recognition*. 2019.

58. Lin, D., et al. *Multi-scale context intertwining for semantic segmentation*. in *Proceedings of the European Conference on Computer Vision (ECCV)*. 2018.

59. Wang, J., et al., *Deep high-resolution representation learning for visual recognition.* IEEE Transactions on Pattern Analysis and Machine Intelligence, 2020.

60. Wolf, T., et al., *HuggingFace's Transformers: State-of-the-art Natural Language Processing.* ArXiv, 2019. **abs/1910.03771**.

61. Khan, S., et al., *Transformers in Vision: A Survey.* ArXiv, 2021. **abs/2101.01169**.

62. Liu, Z., et al., *Swin transformer: Hierarchical vision transformer using shifted windows.* arXiv preprint arXiv:2103.14030, 2021.

63. Kolesnikov, A., et al., *Big transfer (bit): General visual representation learning.* arXiv preprint arXiv:1912.11370, 2019.

64. Jadon, S., *A survey of loss functions for semantic segmentation.* 2020 IEEE Conference on Computational Intelligence in Bioinformatics and Computational Biology (CIBCB), 2020: p. 1-7.

65. Wang, Q., et al., *A comprehensive survey of loss functions in machine learning.* Annals of Data Science, 2020.

66. Sun, R., *Optimization for deep learning: theory and algorithms.* ArXiv, 2019. **abs/1912.08957**.

67. Le, Q.V., et al. *On optimization methods for deep learning*. in *ICML*. 2011.

68. Yang, L. and A. Shami, *On Hyperparameter Optimization of Machine Learning Algorithms: Theory and Practice.* Neurocomputing, 2020. **415**: p. 295-316.

69. Yu, T. and H. Zhu, *Hyper-Parameter Optimization: A Review of Algorithms and Applications.* ArXiv, 2020. **abs/2003.05689**.

70. Xing, F., et al., *Deep learning in microscopy image analysis: A survey.* IEEE Transactions on Neural Networks and Learning Systems, 2018. **29**(10): p. 4550-4568.

71.  Gupta, A., et al., *Deep Learning in Image Cytometry: A Review.* Cytometry Part A, 2019. **95**(4): p. 366-380.

72.  Moen, E., et al., *Deep learning for cellular image analysis.* Nature Methods, 2019. **16**(12): p. 1233-1246.

73.  Belthangady, C. and L.A. Royer, *Applications, promises, and pitfalls of deep learning for fluorescence image reconstruction.* Nature Methods, 2019. **16**(12): p. 1215-1225.

74.  Feng, Y., et al., *Quality Assessment of Synthetic Fluorescence Microscopy Images for Image Segmentation.* 2019. p. 814-818.

75.  Zhang, Y., et al., *A Poisson-Gaussian Denoising Dataset With Real Fluorescence Microscopy Images.* 2019 IEEE/CVF Conference on Computer Vision and Pattern Recognition (CVPR), 2019: p. 11702-11710.

76.  Laissue, P.P., et al., *Assessing phototoxicity in live fluorescence imaging.* Nature Methods, 2017. **14**(7): p. 657-661.

77.  Farsiu, S., et al., *Advances and challenges in super-resolution.* International Journal of Imaging Systems and Technology, 2004. **14**(2): p. 47-57.

78.  van Ouwerkerk, J.D., *Image super-resolution survey.* Image and Vision Computing, 2006. **24**(10): p. 1039-1052.

79.  Yang, J. and T. Huang, *Image super-resolution*, in *Historical Overview and Future Challenges*. 2017, CRC Press. p. 1-33.

80.  Anwar, S., S. Khan, and N. Barnes, *A deep journey into super-resolution.* ACM Computing Surveys, 2020. **53**: p. 1 - 34.

81.  Wang, Z., J. Chen, and S.C.H. Hoi, *Deep learning for image super-resolution: A survey.* IEEE Transactions on Pattern Analysis and Machine Intelligence, 2020: p. 1-1.

82.  Yang, W.M., et al., *Deep learning for single image super-resolution: a brief review.* IEEE Transactions on Multimedia, 2019. **21**(12): p. 3106-3121.

83.  Dong, C., et al., *Image Super-Resolution Using Deep Convolutional Networks.* Ieee Transactions on Pattern Analysis and Machine Intelligence, 2016. **38**(2): p. 295-307.

84.  Lim, B., et al., *Enhanced deep residual networks for single image super-resolution.* 2017 IEEE Conference on Computer Vision and Pattern Recognition Workshops, 2017: p. 1132-1140.

85.  Fang, L.J., et al., *Deep learning-based point-scanning super-resolution imaging.* Nature Methods, 2021. **18**(4): p. 406-416.

86. Wang, H.D., et al., *Deep learning enables cross-modality super-resolution in fluorescence microscopy.* Nature Methods, 2019. **16**(1): p. 103-110.

87. Nehme, E., et al., *DeepSTORM3D: dense 3D localization microscopy and PSF design by deep learning.* Nature Methods, 2020. **17**(7): p. 734-740.

88. Nehme, E., et al., *Deep-STORM: super-resolution single-molecule microscopy by deep learning.* Optica, 2018. **5**(4): p. 458-464.

89. Zelger, P., et al., *Three-dimensional localization microscopy using deep learning.* Optics Express, 2018. **26**(25): p. 33166-33179.

90. Zhang, P., et al., *Analyzing complex single-molecule emission patterns with deep learning.* Nature Methods, 2018. **15**(11): p. 913-916.

91. Qiao, C., et al., *Evaluation and development of deep neural networks for image super-resolution in optical microscopy.* Nature Methods, 2021. **18**(2): p. 194-202.

92. Jones, S.A., et al., *Fast, three-dimensional super-resolution imaging of live cells.* Nature Methods, 2011. **8**(6): p. 499-505.

93. Huang, F., et al., *Video-rate nanoscopy using sCMOS camera–specific single-molecule localization algorithms.* Nature Methods, 2013. **10**(7): p. 653-658.

94. Lin, Y., et al., *Quantifying and optimizing single-molecule switching nanoscopy at high speeds.* PLOS ONE, 2015. **10**(5): p. e0128135.

95. Chen, B., et al., *STORM imaging of mitochondrial dynamics using a vicinal-dithiol-proteins-targeted probe.* Biomaterials, 2020. **243**: p. 119938.

96. Gu, L., et al., *High-Density 3D Single Molecular Analysis Based on Compressed Sensing.* Biophysical Journal, 2014. **106**(11): p. 2443-2449.

97. Ovesný, M., et al., *High density 3D localization microscopy using sparse support recovery.* Optics Express, 2014. **22**(25): p. 31263-31276.

98. Ouyang, W., et al., *Deep learning massively accelerates super-resolution localization microscopy.* Nature Biotechnology, 2018. **36**(5): p. 460-468.

99. Gaire, S.K., et al., *Accelerating multicolor spectroscopic single-molecule localization microscopy using deep learning.* Biomedical Optics Express, 2020. **11**(5): p. 2705-2721.

100. Wagner, N., et al., *Deep learning-enhanced light-field imaging with continuous validation.* Nature Methods, 2021. **18**(5): p. 557-563.

101. Buades, A., B. Coll, and J. Morel. *A non-local algorithm for image denoising*. in *2005 IEEE Conference on Computer Vision and Pattern Recognition (CVPR'05)*. 2005.

102. Boulanger, J., et al., *Patch-based nonlocal functional for denoising fluorescence microscopy image sequences.* IEEE Transactions on Medical Imaging, 2010. **29**(2): p. 442-454.

103. Luisier, F., T. Blu, and M. Unser, *Image denoising in mixed Poisson–Gaussian noise.* IEEE Transactions on Image Processing, 2011. **20**(3): p. 696-708.

104. Weigert, M., et al., *Content-aware image restoration: pushing the limits of fluorescence microscopy.* Nature Methods, 2018. **15**(12): p. 1090-1097.

105. Jin, L., et al., *Deep learning enables structured illumination microscopy with low light levels and enhanced speed.* Nature Communications, 2020. **11**(1): p. 1934.

106. Li, M., et al., *Deep adversarial network for super stimulated emission depletion imaging.* Journal of Nanophotonics, 2020. **14**: p. 016009 - 016009.

107. Möckl, L., A.R. Roy, and W.E. Moerner, *Deep learning in single-molecule microscopy: fundamentals, caveats, and recent developments [Invited].* Biomedical Optics Express, 2020. **11**(3): p. 1633-1661.

108. Hershko, E., et al., *Multicolor localization microscopy and point-spread-function engineering by deep learning.* Optics Express, 2019. **27**(5): p. 6158-6183.

109. Wu, Y., et al., *Three-dimensional virtual refocusing of fluorescence microscopy images using deep learning.* Nature Methods, 2019. **16**(12): p. 1323-1331.

110. Cheng, Y., et al., *A Survey of Model Compression and Acceleration for Deep Neural Networks.* ArXiv, 2017. **abs/1710.09282**.

111. Brown, T., et al., *Language Models are Few-Shot Learners.* ArXiv, 2020. **abs/2005.14165**.

112. Lu, Z., et al. *The Expressive Power of Neural Networks: A View from the Width*. in *NIPS*. 2017.

113. Leshno, M., et al., *Multilayer feedforward networks with a nonpolynomial activation function can approximate any function.* Neural Networks, 1993. **6**(6): p. 861-867.

114. Hoffman, D.P., I. Slavitt, and C.A. Fitzpatrick, *The promise and peril of deep learning in microscopy.* Nature Methods, 2021. **18**(2): p. 131-132.

115. Culley, S., et al., *Quantitative mapping and minimization of super-resolution optical imaging artifacts.* Nature Methods, 2018. **15**(4): p. 263-266.

116. Caicedo, J.C., et al., *Nucleus segmentation across imaging experiments: the 2018 Data Science Bowl.* Nature Methods, 2019. **16**(12): p. 1247-1253.

117. Chai, X., Q. Ba, and G. Yang, *Characterizing robustness and sensitivity of convolutional neural networks for quantitative analysis of mitochondrial morphology.* Quantitative Biology, 2018. **6**(4): p. 344-358.

118. Zhang, Q.-S. and S.-C. Zhu, *Visual interpretability for deep learning: a survey.* Frontiers of Information Technology & Electronic Engineering, 2018. **19**(1): p. 27-39.

119. He, F. and D. Tao, *Recent advances in deep learning theory.* ArXiv, 2020. **abs/2012.10931**.

120. McCann, M.T., K.H. Jin, and M. Unser, *Convolutional Neural Networks for Inverse Problems in Imaging: A Review.* IEEE Signal Processing Magazine, 2017. **34**(6): p. 85-95.

121. Tian, Y., et al., *Superresolution characterization of core centriole architecture.* Journal of Cell Biology, 2021. **220**(4).

122. Huang, Y., et al., *Visualization of Protein Sorting at the Trans-Golgi Network and Endosomes Through Super-Resolution Imaging.* Frontiers in Cell and Developmental Biology, 2019. **7**(181).